# Tantala Kerr-nonlinear integrated photonics

Hojoong Jung[1,2,3], Su-Peng Yu[1,2], David R. Carlson[1], Tara E. Drake[1], Travis Briles[1,2], and Scott B. Papp[1,2,*]

[1]Time and Frequency Division, National Institute of Standards and Technology, 325 Broadway, Boulder, CO, USA
[2]Department of Physics, University of Colorado, 390 UCB, Boulder, CO 80309, USA
[3]Present Address: Center for Quantum Information, Korea Institute of Science and Technology, Seoul, South Korea
*Correspondence to: scott.papp@nist.gov

## Abstract

**Integrated photonics plays a central role in modern science and technology, enabling experiments from nonlinear science to quantum information, ultraprecise measurements and sensing, and advanced applications like data communication and signal processing. Optical materials with favorable properties are essential for nanofabrication of integrated-photonics devices. Here we describe a material for integrated nonlinear photonics, tantalum pentoxide ($Ta_2O_5$, hereafter tantala), which offers low intrinsic material stress, low optical loss, and efficient access to Kerr-nonlinear processes. We utilize >800 nm thick tantala films deposited via ion-beam sputtering on oxidized silicon wafers. The tantala films contain a low residual tensile stress of 38 MPa, and they offer a Kerr index $n_2 = 6.2(23) \times 10^{-19}$ m²/W, which is approximately a factor of three higher than silicon nitride. We fabricate integrated nonlinear resonators and waveguides without the cracking challenges that are prevalent in stoichiometric silicon nitride. The tantala resonators feature an optical quality factor up to $3.8 \times 10^{6}$, which enables us to generate ultrabroad-bandwidth Kerr-soliton frequency combs with low threshold power. Moreover, tantala waveguides enable supercontinuum generation across the near-infrared from low-energy, ultrafast seed pulses. Our work introduces a versatile material platform for integrated, low-loss nanophotonics that can be broadly applied and enable heterogeneous integration.**

Nonlinear interactions of light and material enable powerful controls for electromagnetic fields. Kerr processes in particular allow wavelength conversion of laser light from one color to another [1] and generation of patterns [2] and solitary excitations [3,4] that have many uses. Key use examples include Kerr-microresonator frequency combs based on dissipative solitons [4], which enable frequency-comb functionalities with integrated photonics [5,6], and efficient supercontinuum generation in photonic waveguides, which extends the capabilities of fiber-based modelocked lasers [7]. Leveraging Kerr nonlinearity generally requires low losses that allow extended light-matter interaction, and realization of nearly arbitrary phase-matching conditions. Nanofabrication powerfully enables nonlinear processes by enhancing intensity with small mode volume and by precise lithographic pattern transfer and heterogeneous combinations of materials that facilitate nonlinear phase-matching. These properties are also enabling for integrated (linear) photonics applications, including microwave photonics [8], high-speed optical communication [9], and heterogeneous electronic-photonic integration of lasers and other devices [10]. Therefore, ideal materials are versatile and robust, straightforwardly allowing heterogeneous integration.

Stoichiometric silicon nitride ($Si_3N_4$, hereafter SiN) obtained via low-pressure chemical vapor deposition is an exemplary material for Kerr-nonlinear photonics, offering exceptionally low optical loss [11], moderately strong Kerr nonlinear coefficient [12], and compatibility with CMOS semiconductor nanofabrication [12]. Still large, intrinsic tensile stress (>1 GPa) is a particular challenge with SiN that leads to formation of cracks in the photonics layer [4]. Crack-mitigation strategies require additional, specialized processing steps, including combinations of thermal cycling [13] and damascene growth [14]. Besides SiN, several other photonic-integration materials have been used for the benchmark application of Kerr-soliton generation, including silica with silicon-nitride waveguides [15], deuterated silicon nitride [16], soliton crystals in Hydex [17], aluminum nitride [18], (aluminum) gallium arsenide [19], lithium niobate [20], and gallium phosphide [21].

We demonstrate low-loss, integrated nonlinear photonics in a new platform– tantalum pentoxide ($Ta_2O_5$, hereafter tantala) –which offers excellent material properties to leverage Kerr nonlinearity. Tantala is a dielectric, CMOS-compatible material that has been used in microelectronics [22]. Tantala deposited by ion-beam sputtering (IBS) has long been used in low-loss, high-reflectivity mirror coatings for various fundamental experiments [23]. Over the past four decades at least, tantala has attracted great attention as a photonic material due to its high refractive index, large bandgap, low stress, low optical loss, low thermo-optic (TO) coefficient, and factor of three higher nonlinearity than SiN [24,25]; the majority of tantala's properties [26–30] are comparable or superior to SiN. Recently, IBS-deposited tantala has been used for photonic waveguide fabrication [27] due to its high transparency. In low-confinement tantala-core waveguides, an ultralow loss of 3 dB/m has been reported [27]. However, high-confinement resonators have only been explored with RF sputtered tantala, and only relatively low quality factor ($Q$) has been reported [24]. Four-wave mixing experiments with high-confinement tantala waveguides and relatively low optical loss (1.5 dB/cm) have been reported [24,31,32].

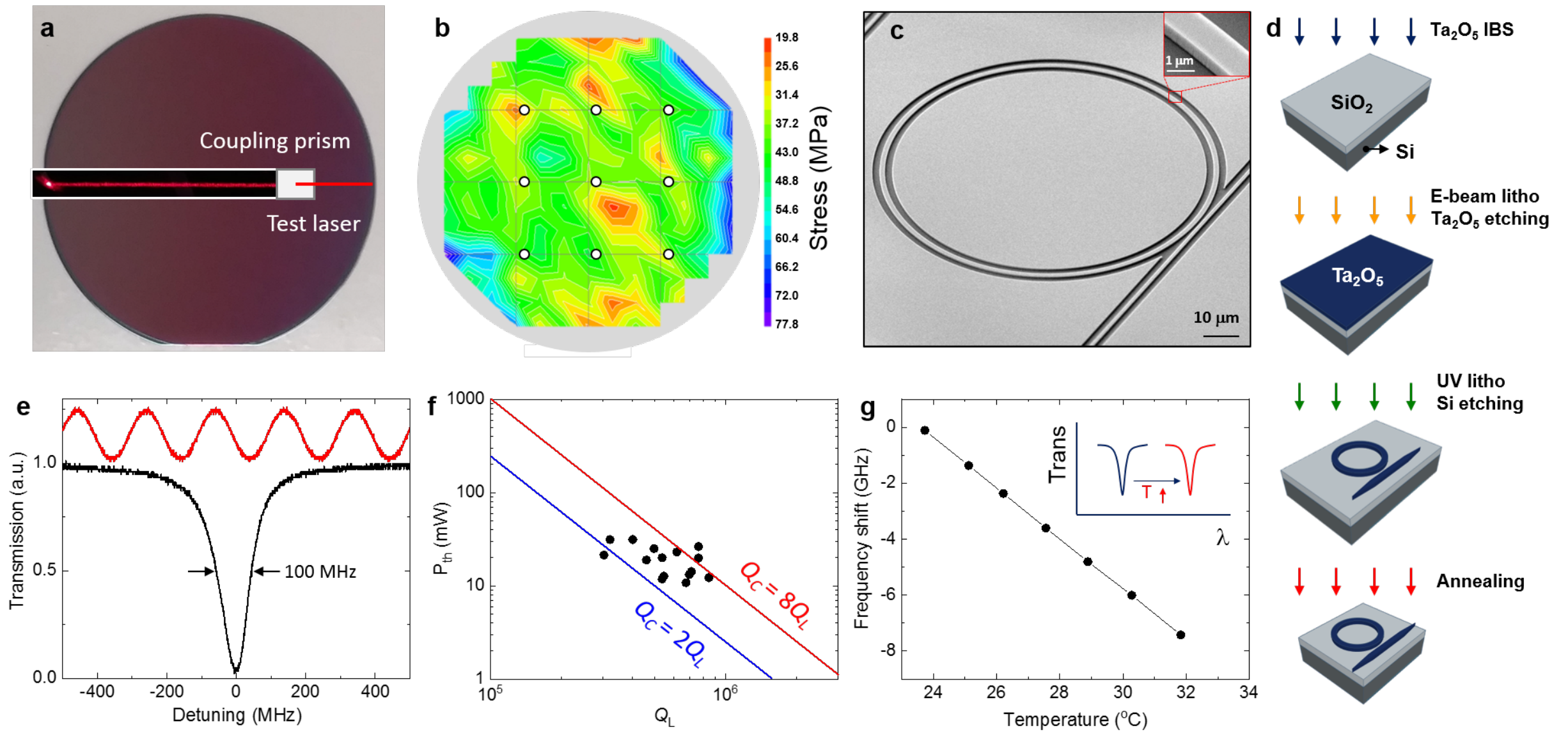

**Fig. 1: Tantala material properties for integrated nonlinear photonics.** (a) Photograph of a crack-free, 820 nm thick tantala layer on a thermally oxidized Si wafer. A laser is prism-coupled to the tantala layer and propagates across the entire wafer. The laser input and prism are shown schematically, and the image indicates propagation. (b) Wafer characterization, indicating a thickness of 817(1) nm at nine points and the tensile-stress map with a mean of 38(7) MPa. (c) Scanning-electron-microscope image of a 46 μm diameter ring resonator. Inset shows the ring waveguide. (d) Fabrication process flow, including IBS deposition, electron-beam lithography, tantala etching, UV lithography, Si etching, and thermal annealing. (e) Resonator lineshape (black trace) with full-width-half-maximum of 100 MHz, calibrated by a 200 MHz free-spectral range fiber Mach-Zehnder interferometer (red trace). (f) Characterization of $n_2$ by $P_{\text{th}}$ for parametric oscillation versus coupling $Q_L$. The blue and red lines indicate the expected change in $P_{\text{th}}$ in the undercoupled regime. (g) Temperature dependence of ring resonator frequency.

In this paper, we report on nanofabrication of high intrinsic $Q_{\text{int}} > 3.8 \times 10^6$ integrated ring resonators and high-confinement waveguides, using tantala photonics layers up to 820 nm thick. The low tensile stress of tantala ensures fabrication of a crack-free photonics layer without the specialized processing required with SiN. We generate Kerr soliton frequency combs with tantala ring resonators, which offer a low parametric oscillation threshold power. Furthermore, we demonstrate supercontinuum generation spanning 1.6 octaves with tantala waveguides with as little as 60 pJ of coupled modelocked laser energy at 1560 nm. We report various tantala material properties, including the $n_2$ and thermo-optic coefficients that we estimate are 6.2(23)×$10^{-19}$ $m^2$/W and 8.8 × $10^{-6}$ /K, respectively. Tantala is a versatile material, offering wide opportunities for heterogenous integration of low-loss and nonlinear photonics due to low-temperature deposition and processing requirements.

**Evaluation of tantala for integrated nonlinear photonics**

Figure 1 explores our tantala material, nanofabrication procedure, and ring resonator characterization. We begin with a tantala film of either 570 nm or 820 nm thickness, which is deposited by way of IBS on a thermally oxidized Si wafer. Two thicknesses are chosen to satisfy two types of phase-matching conditions, one for a top air-cladded waveguide (570 nm) and another for a top silica-cladded waveguide (820 nm). The tantala-IBS service is provided by FiveNine Optics, which offers tantala and $SiO_2$ Bragg reflectors with part-per-million losses [23]. Tantala thickness to >1 μm seems to be feasible. We characterize the material to understand the potential for high-yield, wafer-scale nanofabrication. First as shown in Fig. 1a, we couple a 633-nm laser (Metricon 2010/M prism coupler) to an unpatterned tantala-on-insulator film to visualize cracking and scattering, which has not been observed in the course of fabricating with dozens of wafers. This technique demonstrates a measurement-limited <0.3 dB/cm scattering and absorption loss. Second, as shown in Fig. 1b, we characterize the tantala thickness and stress distributions across a wafer, using the prism coupler for thickness and a multi-beam laser system for stress, respectively. Nine points (open circles in Fig. 1b) indicate a highly uniform film thickness of 817(1) nm. Typically, tantala films are compressively stressed [29], and our as-sputtered film also shows a compressive stress of 170-210 MPa. Thermal annealing to 600 °C, which we also use to reduce optical losses (see below), results in a low tensile stress of 38(7) MPa in the central portion of the wafer. The color map in Fig. 1b shows the measured stress profile with slightly higher values only at the edges due to the wafer fixture.

With tantala-on-insulator wafers we fabricate integrated nonlinear photonics, such as ring resonators; see Fig. 1c. A typical device is 46 μm in diameter, and the ring waveguide width (RW) is 1.5 μm. Our fabrication process (Fig. 1d) involves electron-beam lithography, using a JEOL JBX-6300 FS system and ZEP-520A resist, and pattern transfer, using fluorine ($CF_4$ + Ar) inductively coupled plasma reactive-ion etching (ICP RIE). Scanning-electron microscope (SEM) images and ring-resonator spectroscopy experiments indicate that our lithography pattern transfers roughness to waveguide structures, and this effect should be addressed to reduce waveguide losses. We optionally create a silica upper cladding, using inductively coupled plasma chemical vapor deposition (ICP CVD). To satisfy four-wave-mixing phase matching at 1550 nm, 800 nm (570 nm) tantala films are appropriate in the case of silica (air) cladding on the top and sides of waveguides. We fabricate numerous nonlinear-photonics designs that are organized into chips, which we dice from the full wafer using UV laser lithography and a Si deep reactive-ion etch. Our experiments have demonstrated that thermal annealing of the tantala film is critical to achieve low optical losses, and we perform thermal annealing in a Tystar temperature-controlled vacuum furnace with nitrogen or oxygen background gas. Other steps in our process flow are consistent with silicon-nitride nanofabrication.

We explore the efficiency of Kerr-nonlinear processes with our fabricated tantala devices. In particular for ring resonators (Fig. 1c), we measure $Q$, the nonlinear refractive index ($n_2$), and the thermo-optic coefficient ($dn/dT$), which are key determinants for Kerr-soliton microcombs. To understand how large a quality factor ($Q$) we can achieve in tantala, we fabricate ring resonators with a RW of 5 μm to reduce the effects of scattering from the etched sidewall of the waveguide [33]. We test with a <100 kHz linewidth, 1550 nm laser, which is calibrated by a fiber Mach-Zehnder interferometer; see Fig. 1e. We determine the loaded $Q_L = 1.9 \times 10^6$, which enables us to estimate the internal $Q_{\text{int}} \approx 3.8 \times 10^6$ based on the transmitted power that is roughly consistent with critical coupling. The corresponding propagation loss is 8 dB/m. To understand the nonlinear refractive index ($n_2$) of tantala, we measure the required on-chip threshold power ($P_{\text{th}}$) to observe parametric oscillation. Measurements with 15 separate ring resonators in which $Q_L$ is varied over a small range are shown in Fig. 1f. Based on knowledge of the ring-resonator mode volume, $Q_L$, and refractive index, we estimate nonlinear refractive index $n_2 = 6.2(23) \times 10^{-19}$ m$^2$/W from calculation of the threshold power [34]. To understand the thermo-optic coefficient (Fig. 1g) of our tantala ring resonators, we monitor the thermo-optic (TO) shift in resonance frequency of a particular mode as we heat the temperature of the entire chip. From the slope of the data, we determine $dn/dT \approx -\frac{n}{\nu}\frac{d\nu}{dT} \approx 8.8 \times 10^{-6}$ $K^{-1}$, where $n$ is the refractive index and $\nu$ is optical frequency.

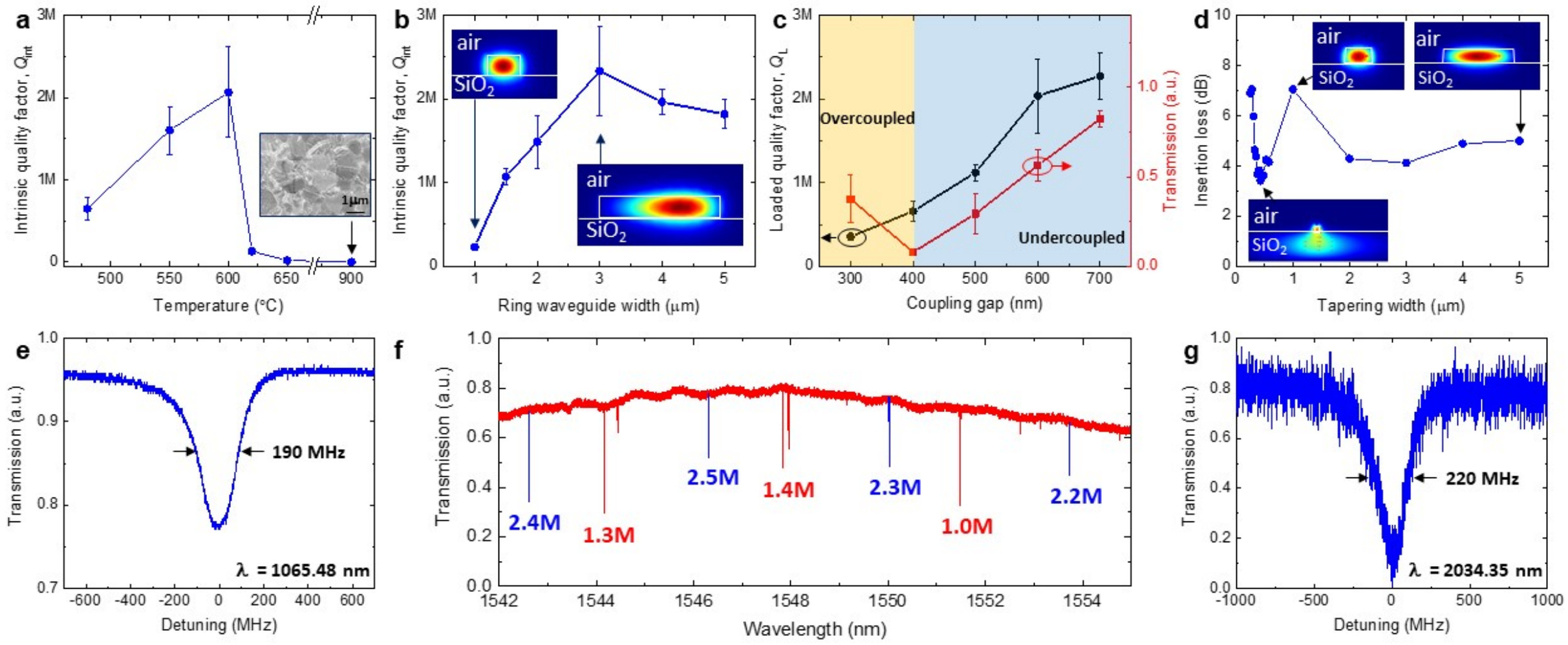

**Fig. 2: Quality-factor measurements of tantala resonators, fabricated with a 570 nm thick film.** (a) $Q_{\text{int}}$ versus annealing temperature of the microring with a width of 3 μm. Inset: SEM image of a presumably crystallized tantala film, following annealing to 900° C. (b) $Q_{\text{int}}$ versus RW. Annealing to 600° for 6 hours was used. Inset: Mode profiles with 1 μm and 3 μm RW. (c) $Q_L$ and normalized transmission versus coupling gap for a 3 μm RW. (d) Edge-coupling insertion loss with a lensed fiber versus inverse-taper width. Inset: Simulated mode profiles for narrow, intermediate, and wide inverse tapers. (e) Resonator lineshape at 1065 nm, indicating $Q_{int} = 1.6 \times 10^6$. (f) Normalized resonator transmission versus wavelength in the C band. Labels indicate $Q_L$ for the TE00 (blue) and TE01 (red) modes with a 600 nm coupling gap. (g) Resonator lineshape at 2034 nm, indicating $Q_{int} = 1 \times 10^6$.

**Quality factor measurements in tantala ring resonators**

Since high $Q_{\text{int}}$ is critical for numerous applications of integrated photonics resonators, in Fig. 2 we explore how the processing and design of tantala devices influences loss. All of the devices considered in Fig. 2 are fabricated with ring resonators of 46 μm diameter, 570 nm tantala thickness, and without a top oxide cladding. As part of our fabrication process flow (Fig. 1d), we anneal the tantala material at relatively high temperature. Figure 2a shows the dependence of $Q_{\text{int}}$ as we vary the maximum annealing temperature from 480 °C to 900 °C; we find little benefit in annealing longer than 5 hours. We observe an improvement in $Q_{\text{int}}$ until 600 °C, especially in the presence of an oxygen background gas that may reduce micro-defects in oxygen-poor IBS tantala films [26,35]. Beyond 600 °C, we presume the tantala film forms a polycrystalline state (inset of Fig. 2a) that does not readily offer low loss [36], however future investigations should consider optimized deposition or post processing to utilize single crystal states [36].

A second challenge in obtaining high $Q_{\text{int}}$ is material scattering and absorption losses, particularly across the broad visible to infrared bands needed for contemporary applications. To characterize scattering loss, we vary RW in fabricated devices from 1 μm to 5 μm and measure $Q_{\text{int}}$; see Fig. 2b. The intrinsic quality factor improves by increasing RW to 3 μm, beyond which our simulations indicate that the mode profile is largely protected from the etched sidewall surface. Simulated TE mode profiles (insets) show that the mode intensity at the waveguide sidewall with RW = 3 μm is less than RW = 1 μm. Moreover, for RW less than 3 μm, we observe resonance-mode splitting characteristic of scattering. We control the resonator coupling rate and chip-edge coupling through e-beam lithography. Figure 2c demonstrates a $Q_L$ tuning across the critical-coupling regime. Figure 2d shows that relatively low-loss edge coupling to lensed fibers with a 2.5 μm mode-field diameter is also possible, using inverse tapers in which the waveguide width is reduced at the chip edge. Even in devices with air cladding on the top and sides, >2 μm wide waveguides tapered at the chip edge provide $\approx -4$ dB per facet due to the large mode size of the wide waveguide. Intermediate-width inverse tapers ≈500-nm wide edge provide $\approx -3$ dB per facet by expanding the mode into the silica cladding. Here, we utilize an inverse-taper length of 20 μm, which causes negligible loss.

Figures 2e-g explore $Q_{\text{int}}$ measurements across the near-infrared wavelength bands of 1–2 μm. We utilize 3 μm RW devices optimized for 1550 nm operation, and we measure $Q_L$ with tunable lasers at 1065 nm (Fig. 2e), the 1550 nm band (Fig. 2f), and 2034 nm (Fig. 2g). Apparently, tantala ring resonators offer high $Q_{\text{int}} > 10^6$ over at least this wavelength range, consistent with the material's reported wide transparency [27]. In particular, across the 1550 nm range, we do not observe wavelength-dependent deterioration of $Q_{\text{int}}$, and we also observe high $Q_{\text{int}}$ in the 1300 nm wavelength band. Our $Q_{\text{int}}$ measurements do not appear to be constrained by tantala absorption; indeed tantala's use in low-loss, bulk reflectors indicates the possibility for $Q_{\text{int}}$ at least twice higher than our present results [37].

**Kerr-soliton frequency-comb generation**

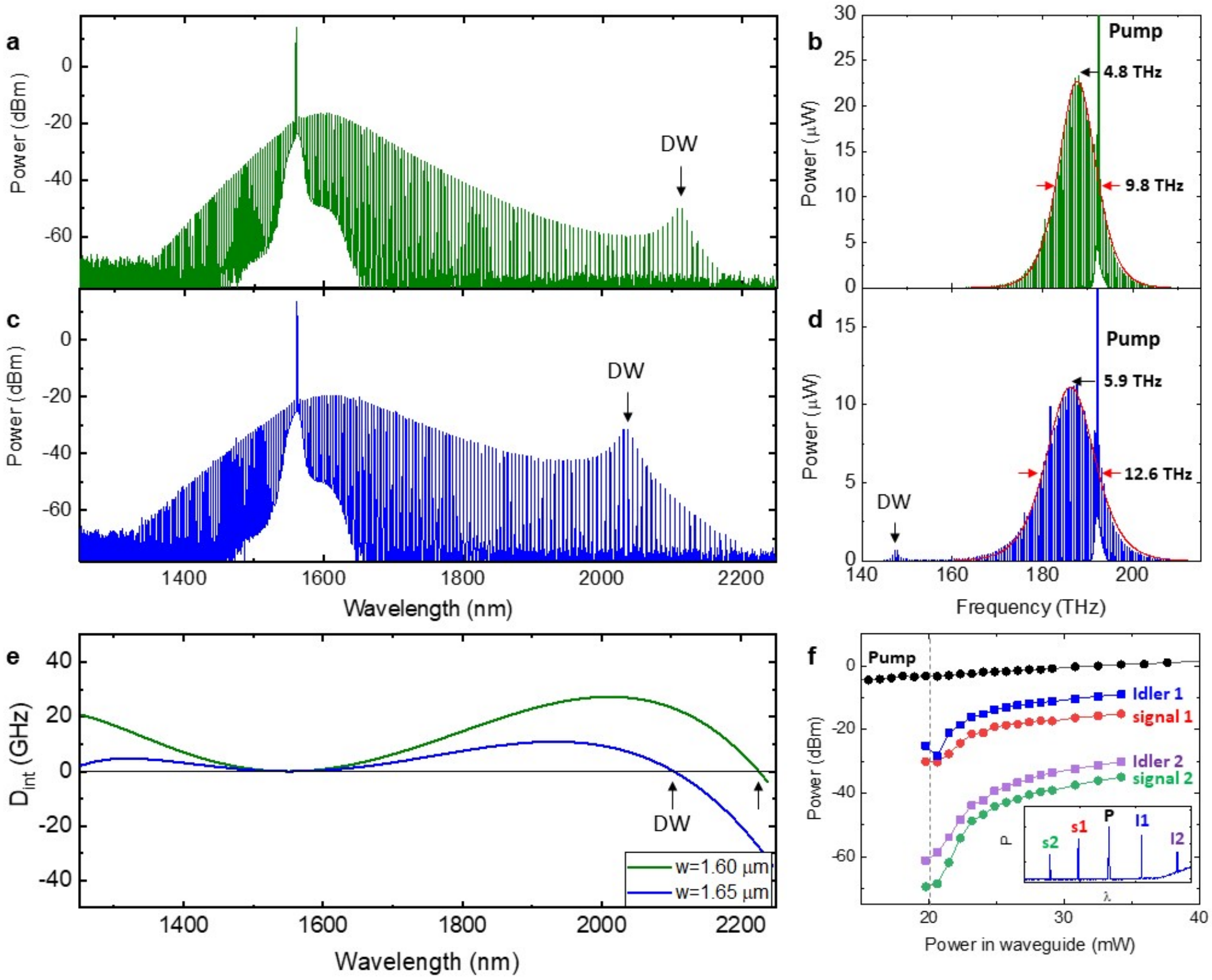

**Fig. 3: Soliton generation in tantala ring resonators.** (a) Observed single-soliton frequency-comb spectrum with a 1.6 μm RW. A dispersive-wave (DW) enhancement is marked. (b) Linear scale plot of the spectrum in (a), highlighting the FWHM bandwidth. (c) Idem of (a) but with a 1.65 μm RW. (d) Linear-scale plot of the spectrum in (b). (e) Calculated integrated dispersion ($D_{\text{int}}$) that characterizes the expected DW wavelength. (f) Power dependence of Turing pattern.

As an application of our high-$Q$ tantala ring resonator platform, we explore Kerr-soliton frequency comb generation; see Fig. 3. Using a tantala device layer thickness of 570 nm, we fabricate ring resonators with 46-μm radius and two settings of RW, ie. 1.6 μm and 1.65 μm, which support ultrabroad-bandwidth anomalous group-velocity dispersion (GVD) and anomalous-to-normal GVD transitions for dispersive-wave engineering [5]. We use a coupling gap of 500 or 600 nm between the ring and a straight waveguide to achieve a slight under-coupling in the 1550 nm band.

We stabilize single soliton pulses in these devices by use of the fast pump-laser-frequency ramping technique [5,34]. Figures 3a,c show the out-coupled spectrum of the soliton in the two RW geometries, including losses from off-chip coupling to lensed fiber. Reduced TO frequency shifts in our tantala resonators facilitate qualitatively easier soliton capture and a reduced requirement for the laser-frequency scan range, compared to our experience with SiN resonators. In Figs. 3b,d, we show a different perspective of the soliton spectra, namely their linear power spectra and comparison to $\text{sech}^2(\nu)$ fit. We estimate a full-width at half-maximum (FWHM) of 9.8 THz (12.6 THz) for the device with RW of 1.6 μm (1.65 μm), corresponding to an output pulse duration of 32 fs (25 fs), respectively. To understand the wavelengths of dispersive-wave generation in our devices, we consider the integrated dispersion (Fig. 3e), based on refractive-index data [38], $D_{\text{int}} \equiv \omega_\mu - (\omega_0 + D_1\mu)$, where $\mu$ is the mode number relative to the pump laser, $\omega_\mu$ is the frequency of mode $\mu$, and $D_1$ is the free spectral range. The dispersive-wave peaks at long wavelength match up with the simulation results. Unfortunately, the wavelength-dependent coupling rate from the ring resonator to a straight waveguide substantially reduces the soliton intensity below ~1400 nm, obscuring the dispersive-wave generation expected around 1100 nm in this device. Pulley couplers with an extended interaction length between the ring and waveguide have been shown to alleviate this issue, and we plan to utilize such designs in future devices [5] From the frequency-comb spectral data, we also estimate the nonlinear refractive index ($n_2$) of our tantala material; see Fig. 3f. Based on the threshold power for optical parametric oscillation, we obtain an estimate of $n_2$. Moreover, we characterize the power contained in the Turing pattern, which initiates with an on-chip, pump-laser power of 20 mW.

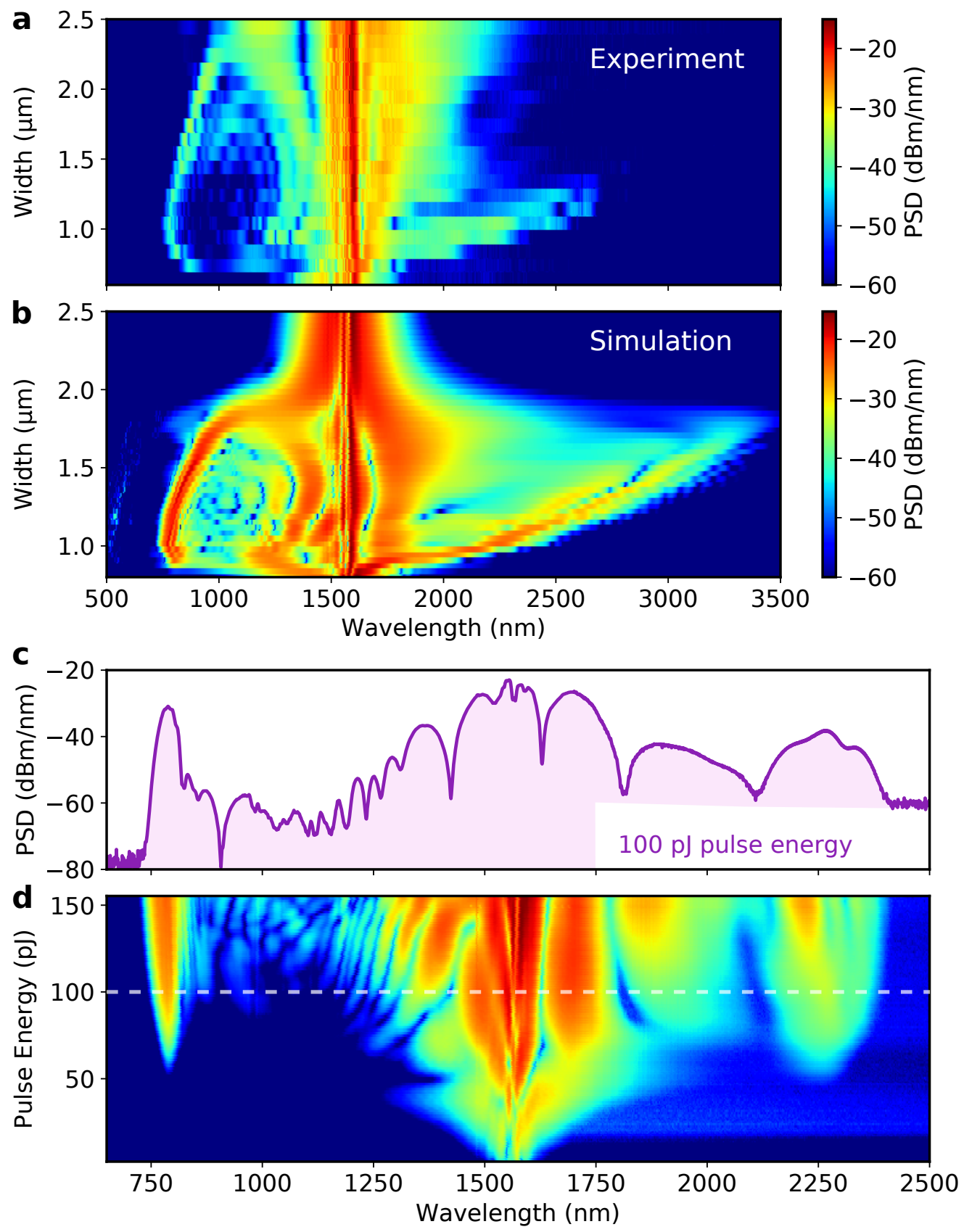

**Fig. 4: Supercontinuum generation in tantala waveguides.** (a) Experimental and (b) simulated output spectra versus WW for 820-nm-thick waveguides with a full $SiO_2$ cladding. We attribute attenuation at wavelengths beyond 2700 nm to absorption in the $SiO_2$ cladding. (c, d) Input power dependence of supercontinuum for a 570-nm-thick, WW=1.5 µm tantala waveguide without $SiO_2$ upper cladding. The trace in (c) corresponds to the white, horizontal line at 100 pJ in (d).

## Supercontinuum generation in tantala waveguides

As a second application of our low-loss tantala integrated-photonics platform, we explore waveguide supercontinuum generation from input modelocked laser pulses. The large $n_2$ index and wide transparency window has the potential to support a variety of supercontinuum device concepts, including low-power frequency combs for optical metrology [39], few-cycle pulse generation [7], and mid-infrared photonics waveguides either via suspended structures [40] or with tantala as a cladding material. These experiments utilize our 820 nm and 570 nm thick tantala films, and an erbium-fiber modelocked laser provides pulses with 80 fs duration and a repetition frequency of 100 MHz [7].

Figures 4a, b explore how GVD engineering controls the supercontinuum spectrum. In particular, we use an 820 nm thick tantala device layer composed of a waveguide, and we apply top and side claddings with ICP-CVD $SiO_2$. To bring 0.5 nJ laser pulses on and off chip, we increase (decrease) the waveguide width (WW) at the chip edge in air-clad (oxide-clad) chips to expand the optical mode in the plane of the chip with minimal wavelength dependence. Therefore, we accomplish input coupling with an aspheric lens and output coupling with lensed fiber at an efficiency of approximately 50% at each facet. The tantala waveguide section for supercontinuum is 5 mm in length, and we vary the GVD in this section by the WW through fabrication of ~20 devices on one chip. We measure the supercontinuum spectrum (Fig. 4a) by coupling the waveguide output to optical-spectrum analyzers that cover 350-5700 nm with an IR compatible multi-mode fiber. Our results indicate that we can create supercontinuum light in the range of 500–2500 nm, with a wide potential for design based on single or multiple WW sections [41]. This closely agrees with our expectation (Fig. 4b), based on finite-element method GVD simulations input to nonlinear Schrodinger equation simulations. The abrupt loss of observed supercontinuum intensity at wavelengths longer than 2500 nm is consistent with O-H absorption in the $SiO_2$ cladding, highlighting the opportunity to develop suspended or ridge-waveguide structures without a second material cladding.

Low-power operation is a key for many supercontinuum applications; therefore, we explore supercontinuum generation in 570 nm thick waveguides without $SiO_2$ cladding that offer the lowest optical loss within our existing nanofabrication process. With air cladding on the top and side, this waveguide thickness is in the optimum range for ultrabroad supercontinuum, particularly f-2f self-referencing with dispersive-wave emission at the second-harmonic wavelength of the input laser. We present a systematic characterization of supercontinuum generation versus on-chip input pulse energy, with Figs. 4c and d showing the spectrum spanning from 750–2400 nm with a 100 pJ input and the spectrum versus pulse energy from 0–150 pJ, respectively. Here, the minimum pulse energy to energize the ~780 nm dispersive wave is 60 pJ, which is comparable to the power needed for supercontinuum in the more mature SiN platform [39]. Future optimization of scattering and absorption losses in tantala waveguides would enable longer length devices and reduced pulse-energy requirements.

**Conclusion**

In summary, we have reported a new integrated-nonlinear-photonics platform based on tantala, which offers ultrabroad-bandwidth Kerr-microresonator solitons and waveguide supercontinuum generation. The key advantages of tantala are: Essentially zero tensile stress even in very thick >800 nm films and <1 nm flatness; a low optical loss <8 dB/m; and a high nonlinear coefficient thrice that of $Si_3N_4$. With these features, we realize a high yield of tantala devices designed to leverage nonlinear optics, without mitigation strategies for film cracking. In the future, we anticipate leveraging tantala's wide transparency window from 320–8000 nm for visible and mid-infrared integrated photonics, which we can access either with $SiO_2$ claddings or without a cladding by way of thick-tantala ridge-waveguide structures or suspended tantala waveguides, respectively. Moreover, the low thermo-optic coefficient of tantala would offer Kerr solitons with reduced fundamental thermo-refractive noise [42].

**Acknowledgements**

This work is supported by NIST and DARPA DODOS and ACES. The authors thank Jennifer Black, Jizhao Zang, and Kartik Srinivasan for helpful comments, and Jeff Chiles and the Boulder Microfabrication Facility. Trade names are provided for information.